\documentclass[nofootinbib,twocolumn,prd,a4paper,final,superscriptaddress,longbibliography]{revtex4-1}
\usepackage{amsmath,amssymb}
\usepackage[top=20truemm,bottom=20truemm, left= 18truemm,right= 18truemm]{geometry}

\usepackage{graphicx}
\usepackage{comment}
\usepackage{bm}
\usepackage{array}
\usepackage[colorlinks=true,allcolors=blue]{hyperref}

\usepackage{color}

\usepackage{setspace}
\usepackage{wrapfig}
\usepackage{subfigure}
\usepackage{here}

\def\pbl{\left(}
\def\pbr{\right)}
\def\sbl{\left[}
\def\sbr{\right]}

\newcommand{\pd}{{\partial}} 

\newcommand{\Real}{\mathrm{Re}} 
\newcommand{\Imag}{\mathrm{Im}}
\newcommand{\Tr}{\mathrm{Tr}} 
\newcommand{\dd}{{\mathrm{d}}}
\newcommand{\CP}{\mathbb{C}P}

\newcommand{\integers}{\mathbb{Z}} 
\newcommand{\reals}{\mathbb{R}} 
\newcommand{\complexes}{\mathbb{C}}

\begin{document}
	\title{Fractional Skyrmion molecules in a \texorpdfstring{$\CP^{N-1}$}{CP(N-1)} model}
	
	\author{Yutaka Akagi}
	\address{Department of Physics, Graduate School of Science, The University of Tokyo, Bunkyo, Tokyo 113-0033, Japan}
	\author{Yuki Amari}
	\address{BLTP, JINR, Dubna 141980, Moscow Region, Russia}
	\address{Department of Mathematical Physics, Toyama Prefectural University, Kurokawa 5180, Imizu, Toyama, 939-0398, Japan}
	\author{Sven Bjarke Gudnason}
	\address{Institute of Contemporary Mathematics, School of Mathematics and Statistics, Henan University, Kaifeng, Henan 475004, P.~R.~China}
	\author{Muneto Nitta}
	\address{Department of Physics, and Research and Education Center for Natural Sciences, Keio University, Hiyoshi 4-1-1, Yokohama, Kanagawa 223-8521, Japan}
	\author{Yakov Shnir}
	\address{BLTP, JINR, Dubna 141980, Moscow Region, Russia}

	\vspace{.5in}
	
	\begin{abstract}
	
	We study fractional Skyrmions in a $\CP^2$ baby Skyrme model with a generalization of the easy-plane potential.
	By numerical methods, 
	we find stable, metastable, and unstable solutions taking the shapes of molecules. Various solutions possess discrete symmetries, and the origin of those symmetries are traced back to congruencies of the fields in homogeneous coordinates on $\CP^2$.
		
	\end{abstract} 
	\maketitle
	
	\section{Introduction}
	\label{sec:intro}

Solitons ubiquitously appear in nature from nonlinear media, water, condensed matter to particle physics, dense stars, and the universe, see e.g.~Refs.~\cite{drazin1989solitons,*babelon}.
	Among solitons, topological solitons are topologically protected stable excitations carrying topological charges, see e.g.~Refs.~\cite{Manton:2004tk,*Shnir2018}. 
	Topological charges are usually integer numbers since they take values in certain homotopy groups which are isomorphic to  integers, ${\mathbb Z}$, or to a discrete group like ${\mathbb Z}_2$.
	
	However, fractionally quantized topological charges may also exist for a variety of reasons.
	An example of a fractional topological charge was first found in the context of vortices.
	The so-called Feynman--Onsager's quantization condition makes
	the circulation of superfluids or magnetic flux of superconductors quantized integers  
	due to the single-valuedness of the wave function, or more mathematically due to fact that they belong to the first homotopy group $\pi_1$.
Nevertheless,
 fractionally quantized vortices are ubiquitously allowed 
  in diverse condensed matter systems with multiple components, 
  such as the A phase of superfluid $^3$He  \cite{Salomaa1985a,*salomaaRMP}, 
 the uniaxially disordered superfluid $^{3}$He~\cite{Autti:2015bta,*Makinen:2018ltj},
unconventional superconductors \cite{Kee2000,*Chung:2007zzc,*Garaud2012a,*Zyuzin2017,*Etter2020,*How2020}, 
multicomponent superconductors \cite{Babaev:2001hv,*Babaev:2004rm,*Babaev:2007,*Goryo2007,*Tanaka2007,*Crisan2007,*Guikema2008,*Garaud:2011zk,*Garaud:2012pn,*Tanaka2017,*Tanaka2018,*Chatterjee:2019jez},
spinor Bose-Einstein condensates (BECs)
\cite{Leonhardt:2000km,*Ho:1998zz,*doi:10.1143/JPSJ.67.1822,*Shinn:2018zde,*Semenoff:2006vv,*Kobayashi:2008pk,*2010arXiv1001.2072K,*Borgh:2016cco},  multicomponent BECs \cite{Son2002,*Kasamatsu2004,*Eto:2011wp,*Cipriani2013,*Tylutki2016,*Kasamatsu2016,*MenciaUranga2018,*Eto2018,*Kobayashi:2018ezm,*Eto:2019uhe,Eto:2012rc,*Eto:2013spa,*Nitta:2013eaa}, 
$^3P_2$ neutron superfluids \cite{Masuda:2016vak,*Masaki:2021hmk}, exciton-polariton condensates \cite{PhysRevLett.99.106401,*Lagoudakis974}, and 
nonlinear optics \cite{PhysRevLett.72.2557,*PhysRevLett.84.634,*Pismen}. 
In high-energy physics, examples can be found in 
quantum chromodynamics at high density \cite{Balachandran:2005ev,*Nakano:2007dr,*Eto:2009kg,*Eto:2009bh,*Eto:2009tr,*Eto:2021nle,*Fujimoto:2020dsa,Eto:2013hoa},  
two-Higgs doublet models which are an important and intensely studied part of physics beyond the Standard Model of elementary particles \cite{Dvali:1993sg,*Eto:2018hhg,*Eto:2018tnk}, 
and supersymmetric gauge theories \cite{Hanany:2003hp,*Auzzi:2003fs,*Eto:2005yh,*Eto:2006cx,*Tong:2005un,*Eto:2006pg,*Shifman:2007ce,*Shifman:2009zz,*Eto:2009bg,*Eto:2009bz,*Gudnason:2010yy}.

Nonlinear sigma (NL$\sigma$) models admit lumps in $d=2+1$ or sigma model instantons in $d=2+0$ (or $d=1+1$) dimensions, 
characterized by the second homotopy group $\pi_2$  \cite{Polyakov:1975yp}.
The gauged $\CP^1$ model (or O(3) model) 
admits fractionally quantized lumps \cite{Schroers:1995he,*Schroers:1996zy,*Baptista:2004rk,*Nitta:2011um};
a single lump is decomposed into two fractional lumps 
which are simultaneously gauged vortices with vortex winding numbers that sum to zero. 
Furthermore, the gauged $\CP^{N-1}$ model, with the maximal torus action $U(1)^{N-1}$ gauged,
admits $1/N$ quantized fractional lumps \cite{Schroers:1995he,*Schroers:1996zy,*Baptista:2004rk,*Nitta:2011um}, 
and 
a single lump is decomposed into $N$ of $1/N$ fractional lumps 
which are simultaneously $U(1)^{N-1}$ gauged vortices with vortex winding numbers that sum to zero. However, no explicit numerical solutions have been constructed for $N>2$ yet and although their fractionality is understood, the interactions between their constituents could be complicated.
Similarly, in the baby Skyrme model, which is the $\CP^1$ model 
with an easy-plane potential and the baby Skyrme term, 
a single baby Skyrmion splits into 
two fractional Skyrmions with global vortex winding numbers that sum to zero 
 \cite{Jaykka:2010bq,*Kobayashi:2013aja,*Kobayashi:2013wra}. 
 In this case, these fractional Skyrmions form a molecule of a certain size determined by the competition of a long-range attraction and a short-range repulsion.\footnote{
 The $\CP^1$ model can be formulated as an Abelian-Higgs model with two complex scalars, in which lumps are replaced by semilocal vortices \cite{Vachaspati:1991dz}. 
 An easy-plane potential, in this case, splits the semilocal vortex into a molecule of two fractional vortices with global vortex charges that sum to zero. This provides a stabilization mechanism of the semilocal vortex which is unstable in the type-II region of parameter space \cite{Eto:2016mqc}.
 }
 Such fractional Skyrmions also appear in chiral magnets 
 described by the $\CP^1$ model with the Dzyaloshinskii-Moriya term
 and an easy-plane potential \cite{Lin:2014ada}, and are sometimes called merons.
 A generalization to Skyrmions in $d=3+1$ characterized by the third homotopy group, $\pi_3$, is also known; in this case, a single Skyrmion is decomposed into two fractional Skyrmions with global monopole charges that sum to zero, forming a molecule \cite{Gudnason:2015nxa}.
 
 The examples described above possess a fractionalization mechanism that is due to the introduction of a suitable potential term.
There are also other mechanisms of fractionalization of topological charges. One mechanism is to  introduce a twisted boundary condition along a compactified direction. A single $SU(N)$ Yang-Mills instanton on ${\mathbb R}^3 \times S^1$ with 
a twisted boundary condition along $S^1$ is decomposed into $N$ fractional instantons with induced monopole charges that sum to zero \cite{Unsal:2007jx}, while 
a single $\CP^{N-1}$ NL$\sigma$ model instanton (Skyrmion instanton in the $SU(N)$ principal chiral model) on ${\mathbb R}^1 \times S^1$ (${\mathbb R}^2 \times S^1$) with 
a twisted boundary condition along $S^1$ is decomposed into $N$ fractional instantons 
with induced domain wall (global vortex) charges that sum to zero  \cite{Eto:2004rz,*Eto:2006mz,*Dunne:2012ae,*Dunne:2012zk,*Misumi:2014jua,*Misumi:2014bsa}
(\cite{Nitta:2015tua}).
Another mechanism is to introduce a suitable modification of the kinetic term. The so-called modified XY model, consisting of a half($1/N$)-periodic nearest-neighbor interaction term in addition to the usual nearest-neighbor interaction term, admits a molecule of two ($N$) half($1/N$)-quantized vortices \cite{Carpenter_1989,*Kobayashi:2019sus,*Kobayashi:2019npl}.

	In this paper, we explore molecules of fractional Skyrmions in a $(2+1)$-dimensional $\CP^{N-1}$ baby Skyrme model.
	We find that fractional Skyrmions with $1/N$ topological charges appear as the minimum constituents and they constitute molecules. In particular, when there are two or more molecules, they form very nontrivial bound states as real molecules of atoms. 
	The Lagrangian density that we consider is given by
	\begin{equation}
	\mathcal{L} = \frac{M^2}{2}\pd_\mu n^a \pd^\mu n^a -\frac{\kappa^2}{4}(f_{abc}n^a\pd_\mu n^b \pd_\nu n^c)^2 - V,
	\label{Lag_n}
	\end{equation} 
	where $M$ and $\kappa$ are real coupling constants, $V$ is a potential term, $f_{abc}$ are the structure constants of $SU(N)$ given by $f_{abc} = -\frac{i}{4}\Tr\pbl\lambda_a\sbl\lambda_b,\lambda_c \sbr \pbr$ with $SU(N)$ generators $\lambda_a$ subject to the normalization $\Tr(\lambda_a\lambda_b)=2\delta_{ab}$.	
	Here, the fields $n^a$ ($a=1,2,..., N^2-1$) satisfy the following $N^2$ constraints
	\begin{equation}
	n^a n^a=\frac{2(N-1)}{N}, \qquad 
	n^a = \frac{N}{2(N-2)}d_{abc}n^b n^c,
	\label{constraints_n}
	\end{equation} 
	where $d_{abc}=\frac{1}{4}\Tr\pbl\lambda_a\{\lambda_b,\lambda_c \} \pbr$, which imply the third-order Casimir identity, $d_{abc}n^an^bn^c = 4(N-1)(N-2)/N^2$. 
	The first term in Eq.~\eqref{Lag_n} is the Lagrangian density of the $\CP^{N-1}$ NL$\sigma$ model, also known as the ($\CP^{N-1}$) Dirichlet term. 
	
	The second term of Eq.~(\ref{Lag_n}) is a Skyrme term introduced as a stabilizer against shrinkage of soliton configurations, and is given by the square of the topological current\footnote{Different types of Skyrme terms have been studied in Refs.~\cite{Ferreira:2010jb,*Ferreira:2011ja,*Amari:2015sva,*Amari:2016ynl}. These Skyrme terms have a rich mathematical structure but are somewhat intricate. Therefore, for simplicity, we will concentrate on the Skyrme term of Eq.~\eqref{Lag_n}.}$^,$ \footnote{$\CP^{2}$ soliton configurations can also be stabilized by the Dzyaloshinskii-Moriya type interaction \cite{Akagi:2021dpk}. }.
	This term can be viewed as the $\CP^{N-1}$ counterpart of the Skyrme term in the $SU(N)/U(1)^{N-1}$ Skyrme-Faddeev model \cite{Faddeev:1998yz,*Amari:2018gbq}, and also a relativistic, $SU(3)$ generalization of the so-called chiral-chiral interaction \cite{grytsiuk2020topological} in the continuum limit.	
	This term has also been used to stabilize a higher-dimensional soliton solution in Ref.~\cite{Radu:2013hca}. 
	
	As for potential terms for $(2+1)$-dimensional $\CP^{N-1}$ models\footnote{In $3+1$-dimensional spacetime, the so-called V-shape potential has also been used to construct compact $Q$-solitons \cite{Klimas:2017eft,*Klimas:2018ywv,*Sawado:2020ncc,*Sawado:2021rsc}.}, a generalization of the easy-axis anisotropic potential, which appears for example in the multi-layered Josephson junction \cite{Nitta:2012xq,*Fujimori:2016tmw}, has been studied in Refs.~\cite{Abraham:1991ki,*Abraham:1992vb,*Gauntlett:2000ib,*Isozumi:2004jc,*Isozumi:2004va,*Eto:2005cp,*Eto:2020cys}.
	Here, we consider instead a generalization of the easy-plane anisotropy term \cite{Jaykka:2010bq}, also known as the XY-type of potential, of the form	
	\begin{equation}
	V=\sum_{l=1}^{N-1} \mu^2 \pbl n^{l(l+2)} \pbr^2,
	\label{pot_N}
	\end{equation}
	where $\mu^2$ is a positive constant.
	This potential is a potential for the field components corresponding to the maximal Abelian torus $U(1)^{N-1}\subset SU(N)$.
	When $N=2$, Eq.~\eqref{pot_N} is just the easy-plane anisotropy term, which generates molecules of half Skyrmions \cite{Jaykka:2010bq,*Kobayashi:2013aja,*Kobayashi:2013wra}.
	For $N=2, 3$, this potential (accompanying the Dirichlet term) appears in a system of BECs with pseudospin-$(N-1)/2$, or in a mixture of $N$ species of bosons \cite{Gra__2014}. We expect that for all $N>1$, the potential emerges from such a pseudospin system.
	This potential is thus a simple and motivated potential giving rise to fractional Skyrmion molecules, but modifications of such type of potential may also possibly give rise to fractional Skyrmions.
	
	In this paper, we numerically construct molecules of fractional Skyrmions with $1\leq Q \leq 4$ in the $N=3$ case, where each constituent possesses a topological charge $1/3$. For $Q=1$, we obtain a stable solution that is a bound state of three fractional Skyrmions forming a triangle, as well as an unstable solution that is a chain of fractional Skyrmions. 
	For $Q=2$, we find a stable solution whose energy density is $\integers_3$ axially symmetric, as well as two unstable solutions composed of fractional Skyrmions that take the shape of a hexagon and a chain, respectively.
	Moreover, we obtain a metastable state which can be regarded either as a bound state of two $Q=1$ stable solutions or as a parallelogram of two $Q=1$ chains. 
	For $Q=3$, we find a stable solution of diamond shape, three metastable states, and two unstable solutions.
	In the $Q=4$ case, we show only a stable solution and a metastable solution, whose energy is quite close to that of the stable solution. In this large topological charge sector, we expect that there exist many metastable and unstable solutions, and we will not attempt to find all of them here.

	This paper is organized as follows. In the next section, we introduce the mathematical structure of the model in detail. In Sec.~\ref{sec:numerics}, we show numerical solutions of molecules of fractional Skyrmions and discuss their properties. Finally, Sec.~\ref{sec:conclusion} concludes the paper with a short discussion.
	
	\section{The model}
	\label{sec:model}

	In this paper, we investigate static solutions of the $\CP^2$ baby Skyrme type model \eqref{Lag_n} with the potential \eqref{pot_N}.
	It follows that the energy of the model is given by $E=E_2+E_4+E_0$ with
	\begin{align}
	&E_2 = \frac{M^2}{2}\int d^2x \pd_i n^a \pd_i n^a,
	\\
	&E_4 = \frac{\kappa^2}{4}\int d^2x (f_{abc}n^a\pd_i n^b \pd_j n^c)^2,
	\\ 
	&E_0 = \mu^2 \int d^2x  \sum_{l=1}^{N-1}  \pbl n^{l(l+2)} \pbr^2,
	\end{align}
	where the subscript $i$ of $E_i$ ($i=2, 4, 0$) indicates the number of derivatives that the functional contains. Since the total energy includes both terms with higher and with lower order of  derivatives than the spatial dimensions (i.e.~the Skyrme term and the potential, respectively), we can clearly evade Derrick's no-go theorem, and Derrick's scaling argument tells us that the solutions should satisfy the scaling relation, $E_4/E_0=1$. 
	
	Let us discuss the (continuous) symmetries of the model. For this purpose, it is convenient to parametrize the field as 
	\begin{equation}
	n^a=-\sqrt{\frac{N-1}{2N}} \Tr\pbl\lambda_aU\lambda_{N^2-1}U^\dagger \pbr,
	\label{n_U}
	\end{equation}
	where $U\in U(N)$.
	When the potential is absent, i.e. $\mu^2=0$, the energy is invariant under the global $U(N)$ transformation $U\to gU$ with $g\in U(N)$, which is equivalent to $n^a \to (n^b/2) \Tr(\lambda_a g\lambda_b g^\dagger)$.
	This symmetry is broken by the potential to $U(1)^{N}$, which is the diagonal subgroup of $U(N)$. 
	In addition, the entire energy functional possesses the hidden local symmetry given by $U\to Uh$ with $h\in SU(N-1)\times U(2)$ commuting with $\lambda_{N^2-1}$, which keeps $n^a$ intact.	

	We are in particular interested in finite energy static solutions. In order for the static energy to be finite, the fields $n^a$ must decay to constant values at spatial infinity (especially for $a=3,8,...,N^2-1$, $n^a$ must vanish at spatial infinity). 
	This condition effectively compactifies physical space, $\reals^2$, to the 2-sphere, i.e.~$\reals^2\cup\{\infty\}\simeq S^2$, and therefore a static configuration with finite energy corresponds to a map from $S^2$ to $\CP^{N-1}$.
	Such configurations are characterized by the topological charge  
	\begin{equation}
	Q=
	\frac{1}{8\pi}\int\dd^2x ~\varepsilon^{jk}f_{abc}n^a\pd_jn^b\pd_kn^c,
	\label{Charge_n}
	\end{equation}
	which is an element of the second homotopy group $\pi_2(\CP^{N-1})=\integers\ni Q$. In this paper, we study fractional Skyrmions, but they always appear as molecules possessing a total topological charge being an integer.

	The fields $n^a$ enable us to interpret the physical meaning of the components in the Lagrangian or energy functional. 
	However, they do not appear to be the best option for constructing solutions because of their complicated nonlinear constraints given in Eq.~\eqref{constraints_n}.
	Thus, we introduce the homogeneous coordinates of $\CP^{N-1}$, $\{Z^\alpha,\bar{Z}^\alpha\}$ ($\alpha=1,2,...,N$) which satisfy the constraint $\sum_{\alpha=1}^{N}|Z^\alpha|^2=1$.
	In terms of the homogeneous coordinates, $n^a$ is given by
	\begin{equation}
	n^a=Z^\dagger \lambda_a Z,
	\label{n_Z}
	\end{equation}
	with $Z=(Z^1, Z^2, ..., Z^N)^{\rm T}$, where T stands for the transpose operation. It follows that $Z$ is an $N$-component complex unit vector, i.e., $Z^\dagger Z=1$.
	One can show that the above parametrization satisfies all the constraints of Eq.~\eqref{constraints_n} by using the identity $(\lambda_a)_{\alpha\beta}(\lambda_a)_{\alpha' \beta'}=2(\delta_{\alpha \beta'}\delta_{\alpha' \beta}- \delta_{\alpha\beta}\delta_{\alpha'\beta'}/N )$.
	This parametrization is equivalent to Eq.~\eqref{n_U} with $U=(Y_1,Y_2,..., Y_{N-1}, Z)$, where $Y_\alpha$ are vectors forming a complete basis in $\complexes^N$ along with $Z$. 
	In terms of $Z$, the energy and the topological charge are given by
	\begin{align}
	E=&\int d^2x \sbl2M^2(D_i Z)^\dagger D_i Z 
	\right.
	\notag\\
	&\qquad\qquad+ \kappa^2 \{(D_i Z)^\dagger D_j Z - (D_j Z)^\dagger D_i Z \}^2 
		\label{Ene_Z}\\
	&\left.\qquad\qquad\qquad\quad
	+ \frac{2\mu^2}{N}\sum_{ \alpha < \beta }\pbl|Z^\alpha|^2-|Z^\beta|^2 \pbr^2 \sbr,
    \notag
	\\
	Q =& -\frac{i}{2\pi}\int d^2x ~\varepsilon^{jk}(D_j Z)^\dagger D_kZ ,
	\label{Charge_Z}
	\end{align}
	where $D_i Z= \pd_i Z - (Z^\dagger\pd_i Z) Z$.
	In the next section, we construct molecules of fractional Skyrmions by optimizing the energy subject to the constraint $Z^\dagger Z=1$.
	
	\begin{figure}[t]
		\begin{center}
			\includegraphics[width=6cm]{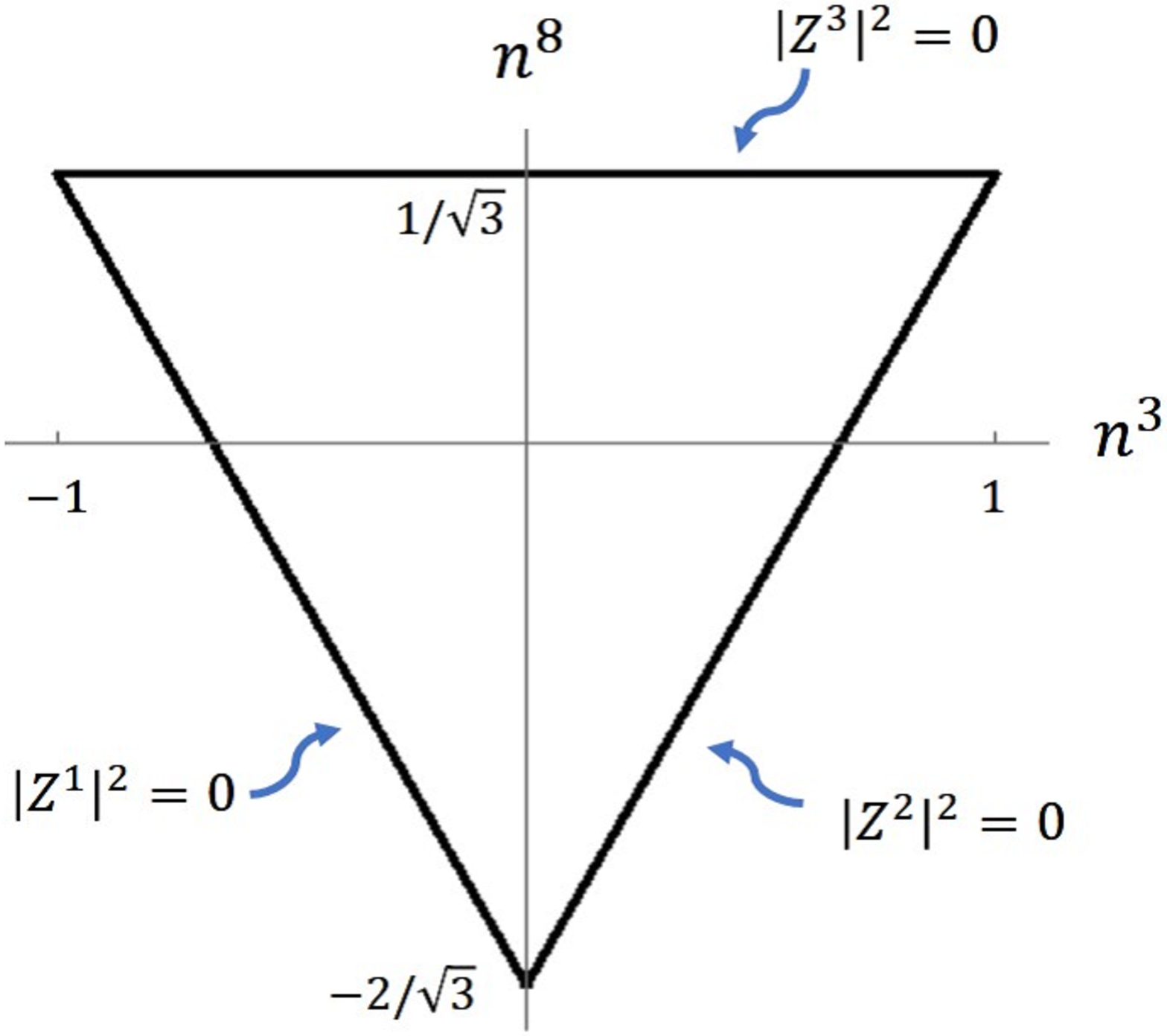}
		\end{center}
		\caption{Toric diagram of $\CP^2$.}
		\label{Fig:toric_diagram}
	\end{figure}

\begin{table*}[t!]
		\centering
		\caption{The numerical values of the energy and topological charge of stable ($\circledcirc$), metastable ($\bigcirc$), and unstable ($\bigtriangleup$) solutions.}
		$\begin{array}{ccccccccc}
		\hline 
		Q &\text{label (shape)}& \text{stability} &E & E_2 & E_4 & E_0 & E_4/E_0 & Q_{\rm Num}  
		\\ 
		\hline\hline
		1 &\text{triangle}&\circledcirc &83.53 &19.58 & 32.07  & 31.88 & 1.006 &  1.000 
		\\ 
		&\text{chain}&\bigtriangleup &84.47 & 21.02 & 31.86  & 31.58 & 1.001 &  1.000 
		\\
		\hline
		2 &\text{curly triangle}&\circledcirc &163.11 & 36.09 & 63.59 & 63.43  & 1.002 & 2.000 
		\\
		&\text{hexagon}& \bigtriangleup&163.14 & 36.56 & 63.35 & 63.24  & 1.002 & 2.000 
		\\ 
		&\text{parallelogram}&\bigcirc &164.28 & 37.01 & 63.80 & 63.47  & 1.005 & 2.000 
		\\ 
		&\text{chain}&\bigtriangleup &167.28 & 40.80 & 63.76 & 62.72  & 1.017 & 2.000 
		\\ 
		\hline
		3 &\text{3A (diamond)}&\circledcirc &243.20 & 51.71 & 95.80 & 95.69  & 1.001 & 3.000 
		\\
		&\text{3B (circle)}&\bigcirc &243.59 & 54.21 & 94.73 & 94.65 & 1.000 & 3.000 
		\\
		&\text{3C}&\bigcirc &243.74 & 52.19  & 95.86 & 95.69 & 1.002 & 3.000 
		\\
		&\text{3D}&\bigcirc &245.05  & 54.02 & 95.68 & 95.35 & 1.003 & 3.000 
		\\
		&\text{3E}& \bigtriangleup&246.41 & 56.55 & 95.03 & 94.83 & 1.002 & 3.000 
		\\
		&\text{3F (chain)}&\bigtriangleup &248.91 & 59.47 & 95.26 & 94.18 & 1.011 & 2.999 
		\\
		\hline
		4	&\text{4A}&\circledcirc &323.27 & 67.61 & 127.92  & 127.75  & 1.001 & 3.999 
		\\
		&\text{4B}&\bigcirc &323.27 & 67.42 & 128.01  & 127.84 & 1.001 & 3.999 
		\\
		\hline 
		\end{array} $
		\label{Table:energy}
	\end{table*}
		
	We introduce a tool to understand the geometrical aspects of the solutions.
	In the idea of toric geometry, the manifold $\CP^{N-1}$ is represented by an $(N-1)$-simplex over which an $(N-1)$-torus $T^{N-1}$ is fibered \cite{Leung:1997tw,Eto:2013hoa}.
	Each $(N-2)$-simplex composing the $(N-1)$-simplex describes $\CP^{N-2}$, and therefore one of the $T^1$ fibers shrink on those.
	At the vertices of the $(N-1)$-simplex, the whole $T^{N-1}$ fiber shrinks.
	Fig. \ref{Fig:toric_diagram} is an example of the representation, known as the toric diagram, for $\CP^2$. On the 
	inside of the triangle, there is a $T^2$ fiber corresponding to the $U(1)^3$ action on the phase of the homogeneous coordinates $Z^\alpha\to e^{i\theta_\alpha}Z^\alpha$ modulo the trivial $U(1)$ action, i.e., $\theta_1=\theta_2=\theta_3$.
	On the edges, one of the $|Z^\alpha|^2$ vanishes, and therefore a $T^1$ fiber shrinks. The fractional Skyrmions we study are interpreted as global vortices whose cores locate where some corresponding fiber shrinks. 
	In addition, on the toric diagram for $\CP^2$, any straight line from a vertex to the opposite edge represents a $\CP^1$ submanifold. If a mapping of the value $(n^3,n^8)$ of a configuration onto the toric diagram describes a straight line, the configuration is a $\CP^1$ embedding in $\CP^2$. On the other hand, when the mapping is two-dimensional, the configuration is a full map to $\CP^2$.

	\section{Numerical solutions}
	\label{sec:numerics}

	We numerically construct molecules of fractional Skyrmions with $1 \leq Q \leq 4$. We restrict ourselves to the $N=3$ case but conjecture that the potential is sufficient for the model to possess fractional Skyrmion molecules for all $N\geq 2$.
	We recognize $Z$ as the fundamental field and perform optimization of the energy under the constraint $Z^\dagger Z=1$.
	Here, the optimization is done by numerically solving  the equations of motion
	\begin{equation}
	\begin{split}
	&\frac{\pd { n^s} }{\pd \Real Z^\alpha}\frac{\delta E}{\delta n^s} - \omega\Real Z^\alpha =0,  
	\\
	&\frac{\pd { n^s} }{\pd \Imag Z^\alpha}\frac{\delta E}{\delta n^s} - \omega\Imag Z^\alpha =0,  
	\end{split}
	\label{eom}
	\end{equation}
	where 
	$\omega$ is a Lagrange multiplier. 
	The functional derivative of $E$ with respect to $n^s$,  $\delta E /\delta n^s$, 
	is written as
	\begin{align}
	    &\frac{\delta E}{\delta n^s} \equiv \frac{\pd {\cal E}}{\pd n^s} - \pd_k \frac{\pd {\cal E}}{\pd (\pd_k n^s)}
	    \notag\\
	    & =-M^2\pd_i^2 n^s 
	    +\frac{\kappa^2}{2}\pbl 2\pd_i F_{ij}R^s_j+3F_{ij}\pd_iR^s_j \pbr
	    \notag\\
	    &\qquad\qquad\qquad\qquad\qquad
	    +2\mu^2(\delta_{s,3} n^3 + \delta_{s,8}n^8),
	\end{align}
	where ${\cal E}$ denotes the energy density and
	\begin{equation}
	    F_{ij}=f_{abc}n^a\pd_in^b\pd_jn^c, \quad R^a_j =f_{abc}n^b\pd_jn^c.
	\end{equation}
	Replacing spatial derivatives by their 4th-order finite difference approximations, we explore the solutions of Eq.~\eqref{eom} using a nonlinear conjugate gradients method, with the augmented Lagrangian method.

	To prepare initial configurations, we write 
	\begin{equation}
	Z=\frac{\pbl1,u_1, u_2\pbr^T}{\sqrt{1+|u_1|^2+|u_2|^2}},
	\end{equation}
	where $u_1$ and $u_2$ are complex scalar fields, or inhomogeneous coordinates of $\CP^2$.
	Finiteness of the energy requires that both $n^3$ and $n^8$ vanish at spatial infinity, which corresponds to $|u_1|^2$ and $|u_2|^2$ both approaching unity. 
	Guided by this condition, we employ initial configurations of a single lump configuration with topological charge $Q$ of the form
	\begin{equation}
	u_1=1+\frac{r_1}{(x+iy)^{Q}}, 
	\qquad
	u_2=1-\frac{r_2}{(x+iy)^{Q}}, 
	\label{ini_single}
	\end{equation}
	where $r_1$ and $r_2$ are nonzero real constants.
	In order to check the stability of solutions obtained using the initial configuration \eqref{ini_single} and to construct metastable solutions that cannot be obtained when we use Eq. \eqref{ini_single},
	we also 
	conduct simulations employing
	initial configurations with $Q$ separated integer Skyrmion states  
	\begin{equation}
	\begin{split}
	 & u_1= 1 + \sum_{k=1}^{Q}\frac{r^{(1)}_{k}e^{i\theta^{(1)}_k}}{x - a_k + i (y - b_k)}, 
	\\
	& u_2= 1 + \sum_{k=1}^{Q}\frac{r^{(2)}_ke^{i\theta^{(2)}_k}}{x - a_k + i (y - b_k)},
	\end{split}
	\label{ini_separated}
	\end{equation}  
	where $r^{(j)}_k$ are positive constants, and $\theta^{(j)}_k$, $a_k$ and $b_k$ are real constants. Since the initial configurations \eqref{ini_single} and \eqref{ini_separated} are both holomorphic functions, they are BPS solutions of the $\CP^2$ NL$\sigma$ model, i.e. in the $\kappa=\mu=0$ case.

    \begin{figure*}[t]
	\begin{center}
		\includegraphics[width=16cm]{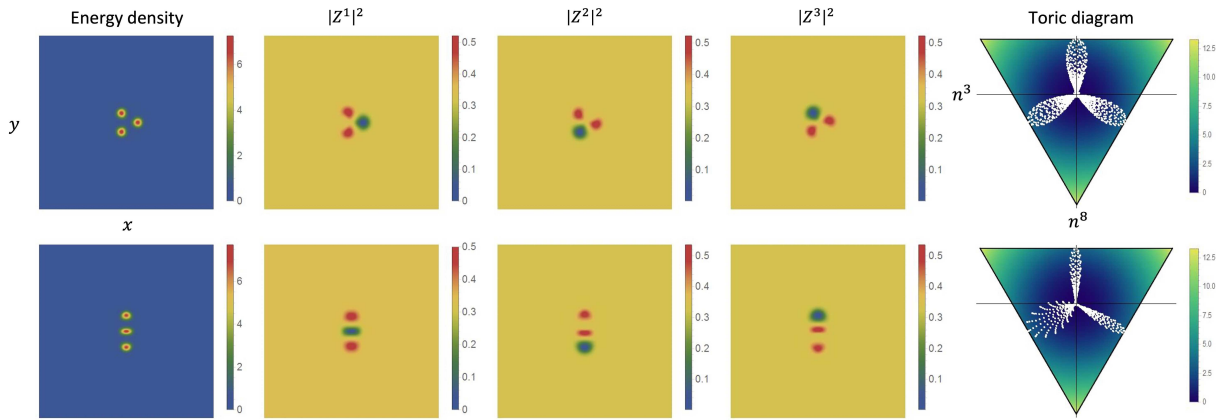}
	\end{center}
	\caption{The energy density, the norm of the homogeneous coordinates, and the toric diagram of the $Q=1$ solutions with the coupling constants $(M^2,\kappa^2, \mu^2) =(1,16,10)$. The dots in the toric diagram represent the values 
	$(n^3,n^8)$ 
	on each lattice point of the numerical solution, and the color of the background shows the value of the potential.}
	\label{Fig:conf_Q=1}
	\end{figure*}

	\begin{figure*}[t]
	\begin{center}
		\includegraphics[width=16cm]{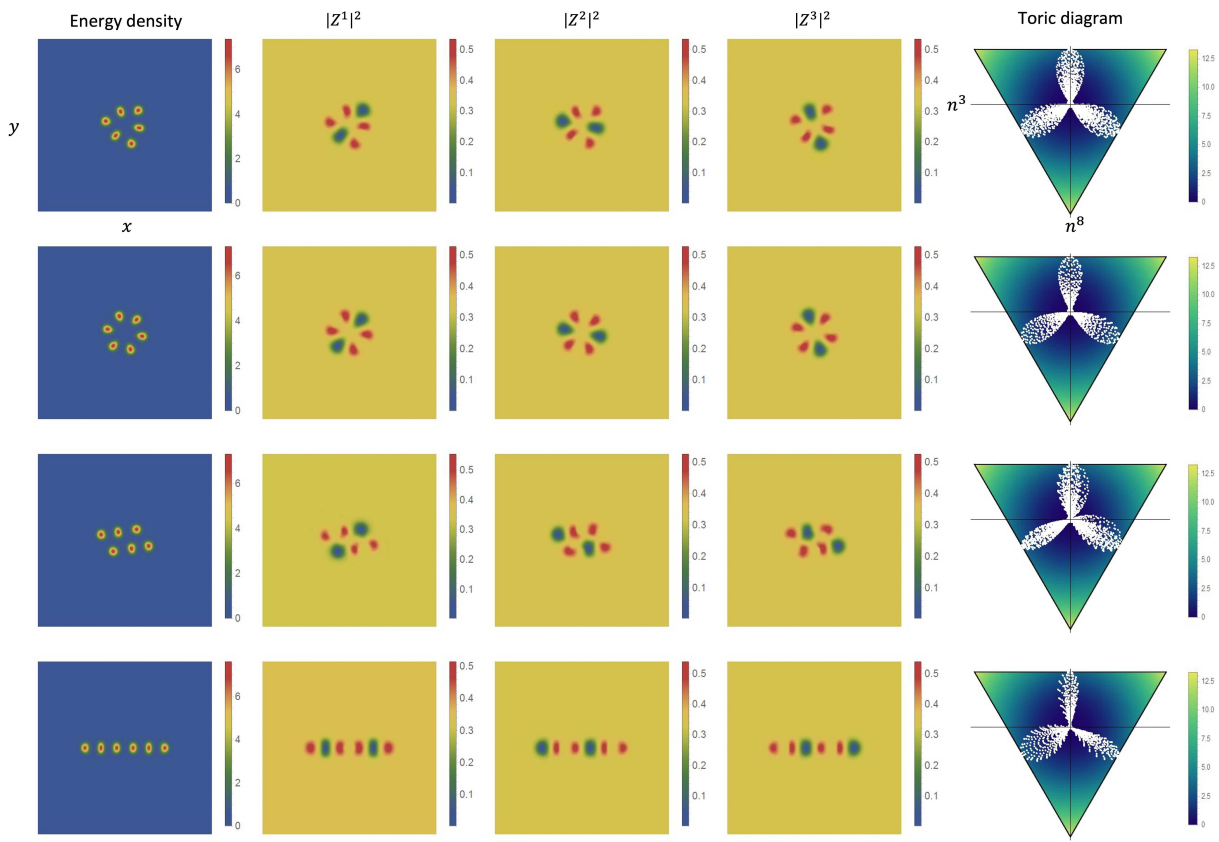}
	\end{center}
	\caption{The energy density, the norm of the homogeneous coordinates, and the toric diagram of the $Q=2$ solutions. For more details, see the caption of Fig.~\ref{Fig:conf_Q=1}.}
	\label{Fig:conf_Q=2}
	\end{figure*}

\begin{figure*}[t]
		\begin{center}
			\includegraphics[width=8.5cm]{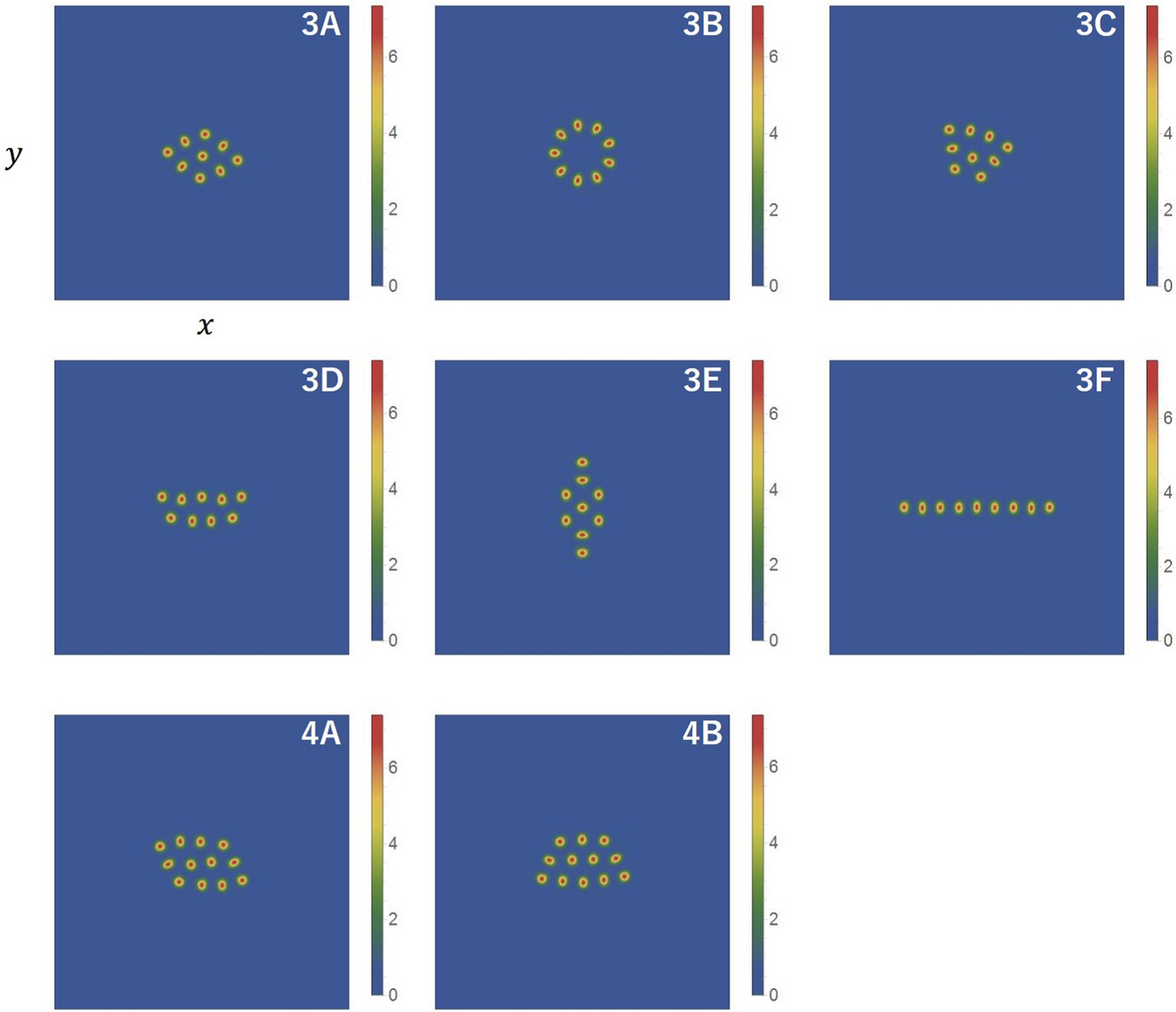}
			\includegraphics[width=8cm]{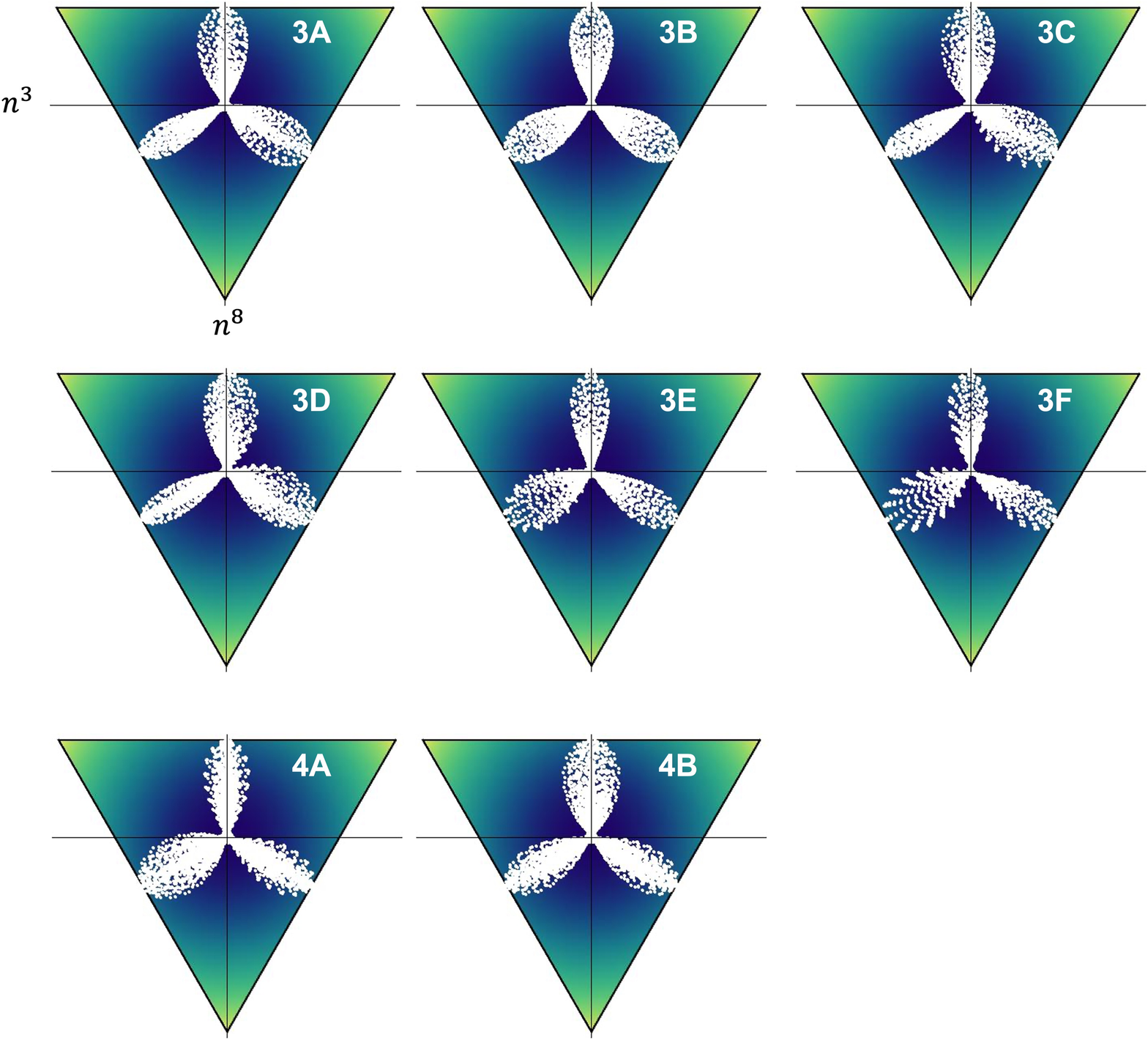}
		\end{center}
		\caption{The energy density (left) and toric diagram (right) of the solutions with $Q=3$ and $Q=4$. }
		\label{Fig:conf_Q=34}
	\end{figure*}

	We performed numerical simulations with $201^2$ lattice sites for $Q=1, 2$ and $301^2$ lattice sites for $Q=3, 4$,
	fixing the lattice spacing to $\Delta = 0.2$ and the coupling constants as $(M^2,\kappa^2, \mu^2) =(1,16,10)$.
	We show the energy and topological charge of the solutions we obtain in Table \ref{Table:energy}. As shown in the table, the numerical values of the topological charge of the solutions, $Q_{\rm Num}$, are close to their desired integer values, and the Derrick scaling relation, i.e.~$E_4/E_0=1$, is satisfied very well too.	
	We show the energy distribution, the norm of the homogeneous coordinates, and a scatter plot of $(n^3,n^8)$ on the toric diagram of the $Q=1$ solutions in Fig.~\ref{Fig:conf_Q=1}. The solutions with $Q=2$ are similarly given in Fig.~\ref{Fig:conf_Q=2}, and for the $Q=3$ and $Q=4$ solutions, we show only the energy densities and the toric diagrams in Fig.~\ref{Fig:conf_Q=34}, due to a large number of solutions.
	The topological charge distribution is similar to that of the energy, so we do not present it to avoid redundancy.
	We will adopt the following nomenclature for describing the stability of the solutions, based on quantum mechanical language:  stable solutions are the global minima of the energy functional in solution space; metastable are local minima and unstable solutions have at least one run-away direction where the solution classically can fall into a lower-energy state. 
	In the following, we discuss the properties of the solutions with each topological charge in order.
	
	For $Q=1$, we find molecules of three fractional Skyrmions of the same size, forming an equilateral triangle, as the stable state. 
	In addition, unlike the $\CP^1$ case \cite{Kobayashi:2013aja} there exists an unstable configuration, a chain of fractional Skyrmions.
	For both solutions, the dots in the toric diagram in Fig.~\ref{Fig:conf_Q=2} scatter like a trefoil, and hence the solutions are genuine $\CP^2$ configurations.
	The symmetry of the trefoils comes from the congruency of the norm of the homogeneous coordinates: the trefoil for the triangle solution has  $\integers_3$ symmetry because all $|Z^\alpha|^2$ are congruent to each other; on the other hand, that of the chain is not $\integers_3$ but $\integers_2$ symmetric, because $|Z^1|^2$ is not congruent to both $|Z^2|^2$ and $|Z^3|^2$.
	As expected from the discussion of the toric diagram, one can observe in Fig.~\ref{Fig:conf_Q=1} that the position of Skyrmions, i.e., the energy peaks coincide with the zeros of the homogeneous coordinates, in the same way as the positions of fractional vortices in multicomponent BECs correspond to zeros of the condensate wave functions \cite{Kobayashi:2013wra,Eto:2012rc,Eto:2013spa,Nitta:2013eaa}.

	For $Q=2$, we obtain several molecules of six fractional Skyrmions as (meta)stable or unstable configurations. Among them, the lowest energy configuration is the curly triangle. The second-lowest is the hexagon, which finally decays into the curly triangle configuration after a large number of iterations and so is unstable. 
	The energy density of the hexagon has literally $\integers_6$ axial symmetry, which stems from the mirror symmetry of $|Z^\alpha|^2$ and the congruence relation on the norm of the homogeneous coordinates.
	The norm of the homogeneous coordinates of the curly triangle type are also 
	congruent with each other, and the congruency provides the $\integers_3$ symmetry of its energy density.
	In addition to these two configurations, we have a metastable solution that is a parallelogram state and an unstable solution, i.e.~a chain of Skyrmions.
	Their energy density is $\integers_2$ axially symmetric.
	One can regard the parallelogram state 
	as a bound state of two $Q=1$ triangles 
	or as a parallelogram of two $Q=1$ chains. 
	On the other hand, the $Q=2$ chain is a linearly connected state of two $Q=1$ chains. 
	We can observe from Fig.~\ref{Fig:conf_Q=2} that two Skyrmions, corresponding to zeros in the same component $|Z^\alpha|^2$, repel each other, and those in different components attract.

	In the $Q=3$ case, the stable solution is a configuration of diamond shape. 
	Like the $Q=2$ case, the most symmetric configuration, a molecule of nine fractional Skyrmions sitting on a circle, has slightly higher energy than the lowest energy state, although this state is not unstable but only metastable.
	Moreover, we find two further metastable and two unstable states.
	
	For $Q=4$, the situation becomes a bit different from the $Q=3$ case; a molecule of fractional Skyrmions sitting on a circle does not appear as a metastable configuration; and there appear two states, 4A (a parallelogram) and 4B (a trapezoid), whose total energies are almost equal to one another.

	As a generalization of the results shown above, it is natural to believe that in general, the potential will allow for molecules of $3Q$ fractional Skyrmions. In addition, we conjecture that the potential \eqref{pot_N} equips the model with solutions which are molecules of $NQ$ factional Skyrmions for $N>3$. 	
	
	\section{Summary and discussion}
	\label{sec:conclusion}
	
In this paper, we have studied fractional Skyrmions in a $\CP^2$ baby Skyrme model made of the standard kinetic term, the Skyrme term generalized to $\CP^2$ and a generalization of the easy-plane potential. The formulation of the model in terms of homogeneous coordinates is the most convenient for numerical calculations, whereas a corresponding inhomogeneous coordinate representation proved useful for generating initial conditions. The solutions we have found are all molecules of $NQ$ (with $N=3$) fractional Skyrmions with each constituent carrying $1/N$ of the topological charge $Q$. Furthermore, all the solutions are genuine $\CP^2$ solutions, which we have shown by generating scatter plots on the toric diagram of $\CP^2$ of the numerical solutions. For $Q=1$, the stable solution is a triangular composition of fractional Skyrmions, and the unstable solution is a chain that eventually will bend over and collapse into the triangle. For $Q=2$, the stable solution has $\integers_3$ symmetry, and we have denoted it as the curly triangle solution. A parallelogram turns out to be a metastable state, whereas a hexagon and a chain made of two $Q=1$ chains are both unstable states. We have also found stable, metastable, and unstable solutions for $Q=3,4$.

While we have studied configurations of only lower $Q$ in this paper, those of higher $Q$ remain as a future problem. 
In particular, revealing the structure at sufficiently large $Q$, is the most important problem.
In the case of the $\CP^1$ model, 
the ground state is a square lattice for sufficiently large $Q$ \cite{Jaykka:2010bq,*Kobayashi:2013aja,*Kobayashi:2013wra}. 
This situation also occurs in chiral magnets with an easy-plane potential \cite{Lin:2014ada}.
We thus expect that in our case of the $\CP^2$ model, the ground state, for sufficiently large $Q$, is a triangular lattice.

A further natural continuation of this study would be to see if the $1/N$ fractional Skyrmions persist for the $\CP^{N-1}$ model with our generalization of the easy-plane potential, as we have conjectured. In particular, it is interesting to determine whether the ground states of $Q=1$ configurations are square, pentagon, hexagon, or more generally polygons 
for $N=4, 5, 6$ and so on. For sufficiently large $Q$, what kind of ground state is realized is a very interesting question to explore. Since a plane can be filled by tiling a square, triangular, or hexagonal lattice, 
this may correspond to $N=2(4), 3$, or $6$, respectively. It is, however, impossible for general $N$ to fill a plane periodically. 
It is also interesting to study fractional Skyrmions in the flag manifold NL$\sigma$ model \cite{Affleck:2021ypq,*Ueda_2016,*Amari:2017qnb} with a Skyrme term and an easy-plane type potential.
It would also be interesting to study generalizations of the potential considered in this paper and classify which types of potentials give rise to fractional solitons. We leave these issues for future studies. 

The fractional Skyrmions studied here may very well exist also in systems directly applicable to condensed matter physics. The most imminent and promising situation is where the model is modified to use a generalization of the Dzyaloshinskii-Moriya term, instead of the Skyrme term. Such a model is expected to be realizable in a mixture of $N$ species of bosons with artificial gauge potentials \cite{Gra__2014}.

Another direction is to investigate whether this model admits stable Hopfions in $d=3+1$. Since Hopfions are topologically unstable 
in $\CP^{N-1}$ models due to $\pi_3(\CP^{N-1})=0$
except for $N=2$, this is a highly nontrivial dynamical question.
If they are stable in a certain parameter region, they would be fractional Hopfion molecules, which generalize the fractional Hopfions in the 
usual Faddeev-Skyrme model with a symmetry-breaking potential \cite{Samoilenka:2017hmn},
possibly of the form of closed fractional lump-strings linking each other.

	\subsection*{Acknowledgments}
	Y.~Akagi thanks Hosho Katsura for a helpful discussion.
	The work of Y.~Akagi is supported by JSPS KAKENHI Grant No.~JP20K14411 and JSPS Grant-in-Aid
for Scientific Research on Innovative Areas ``Quantum Liquid Crystals'' (KAKENHI Grant No.~JP20H05154).
	S.~B.~G.~thanks the Outstanding Talent Program of Henan University for
partial support.
The work of S.~B.~G.~is supported by the National Natural Science
Foundation of China (Grants No.~11675223 and No.~12071111).
The work of M.N.~is supported in part by JSPS Grant-in-Aid for
Scientific Research (KAKENHI Grant No.~18H01217).
Ya.~S.~gratefully acknowledges support by the Ministry of Education of Russian Federation, project FEWF-
2020-0003. The computations in this paper were run on the "GOVORUN" cluster supported by the LIT, JINR.
	
 \bibliographystyle{apsrev4-1}	\bibliography{reference}
	
\end{document}